# Graphene-based Analytical Lab-on-Chip Devices for Detection of Viruses: A Review

*Joydip Sengupta[1], Arpita Adhikari[2] and Chaudhery Mustansar Hussain[3]\**

[1]Department of Electronic Science, Jogesh Chandra Chaudhuri College, Kolkata - 700033, India

[2]Department of Electronics and Communication Engineering, Techno Main Salt Lake, Kolkata-700091, India

[3]Department of Chemistry and Environmental Science, New Jersey Institute of Technology, USA

*Corresponding author Email: chaudhery.m.hussain@njit.edu

**Abstract:**

Lab on a chip (LOC) device intakes fluid and makes it flow through the microchannels, to achieve rapid, highly sensitive and low-cost analysis with significant yield. Graphene has vast potential to be used in LOC devices owing to its remarkable and unique properties. A trustworthy, swift, inexpensive and facile detection scheme is of urgency due to the current situation of COVID-19 to break the chain of transmission and lab-on-chip based biosensor has materialized itself as a realistic solution for this purpose. The addition of graphene has augmented the sensing capability of the LOC devices to a superior level. Recently, graphene-based lab-on-chip type biosensor is effectively used for faithful detection of SARS-CoV-2 and this draws the attention of the researchers to review the recent progress in graphene-based LOC platforms for detection of viruses and the same has been reviewed here.

## 1. Lab-on-a-chip: Materials and Fabrication

"Miniaturization" has created a revolution in electronic devices by downsizing them from microelectronics to nanoelectronics. Primarily, Moser et al.[1] coined the term "Lab on Chip" to describe the miniaturized thin-film biosensors build-up by them. Afterwards, the term LOC symbolizes a microfluidics technology-based miniaturized device that can be scaled down and also able to assimilate various laboratory operations onto a structure that can be fitted on a chip.

Various types of inorganic, organic and composite materials are employed to fabricate basic structure LOC devices. Among semiconductors, silicon is employed as the base material for LOC[2] owing to its good chemical stability, structural tunability and medical compatibility. To cut off the large production cost of Si, glass is also employed to build LOC devices having the additional advantage of optical transparency over Si. Among the other materials, low-temperature cofired ceramic (LTCC)[3] is also used. Various types of organic materials are often employed to fabricate LOC devices such as polymer, paper and hydrogel. Among polymers, two kinds of polymers namely, elastomers and thermoplastic are used. As the performance of the LOC devices largely depends on the fabricated structure, thus, numerous aspects of LOC, along with the required fabrication material was reviewed[4] and the results indicated that the inclusion of nanomaterials will enhance the performance of the LOC[5] irrespective of the material used for structural development. Consequently, various nanomaterials were used in LOC[6] for performance improvement, however, graphene becomes peerless in terms of its inclusion in LOC[7] owing to its novel structure and exotic physio-chemical characteristics.

Regarding the fabrication techniques, both top-down and bottom-up methods are employed. Alongside top-down or bottom-up, many sub-processes such as etching, lithography, thermoforming, molding, hot embossing, polymer ablation, polymer casting, bonding are also involved in the final fabrication of the LOC device (Fig 1).

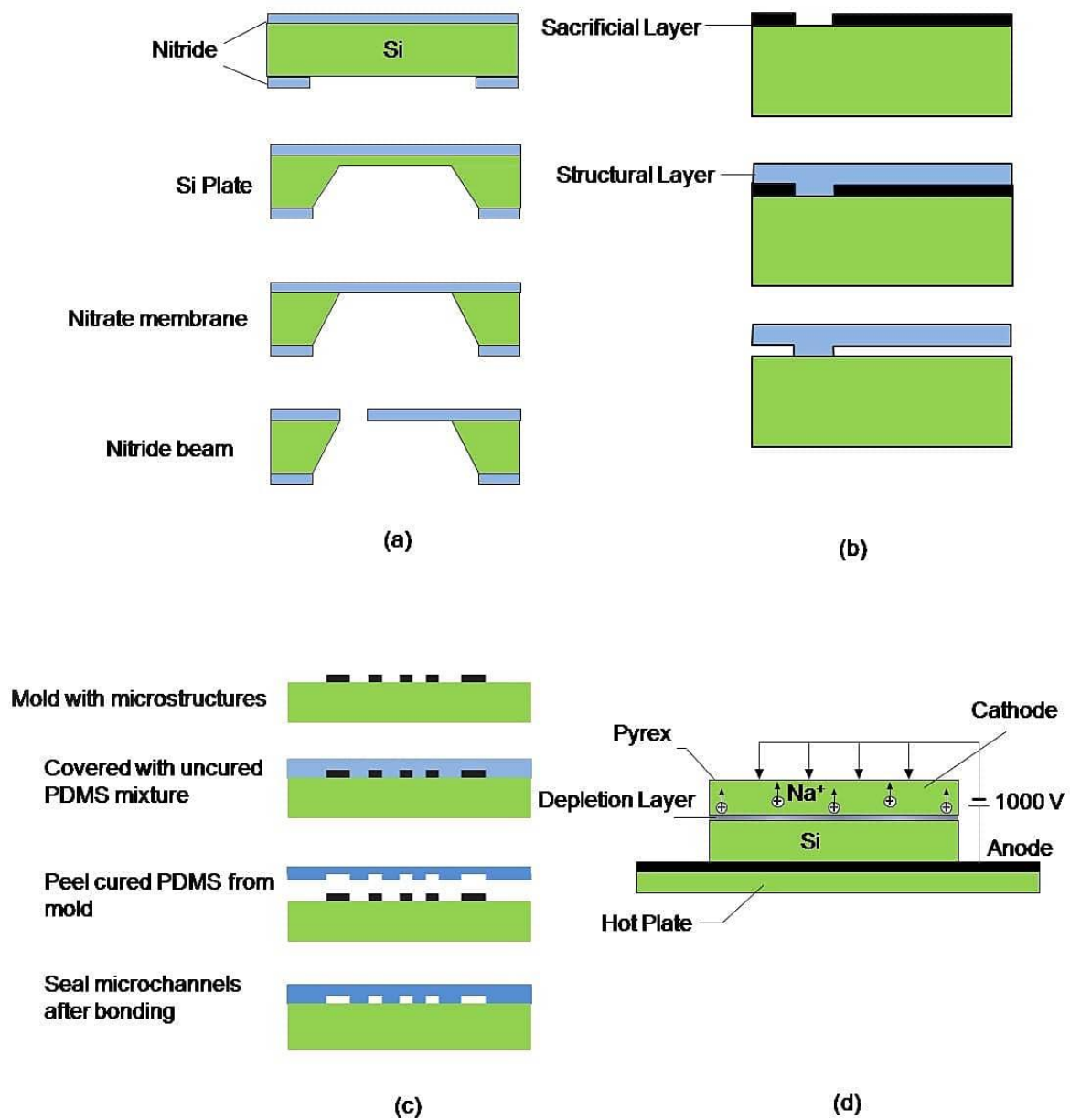

*Fig 1. Different routes LOC fabrication: a) Bulk Micromachining, b) Surface Micromachining, c) SoftLithography, d) Bonding[8].*

## 2. Graphene: Synthesis and application as a lab-on-chip material

The production procedure and performance of the LOC device are governed by microfluidic technology and the raw material required for the fabrication of such devices should possess some distinct characteristics[9]. In addition, it was also suggested that nanomaterials such as graphene will be the preferred candidate for LOC device fabrication[10]. The honeycomb structure of ultra-lightweight graphene has novel chemical and physical properties[11] likewise extreme optical transparency, outstanding mechanical properties, exceptionally high thermal conductivity, superior electron mobility etc (Fig 2).

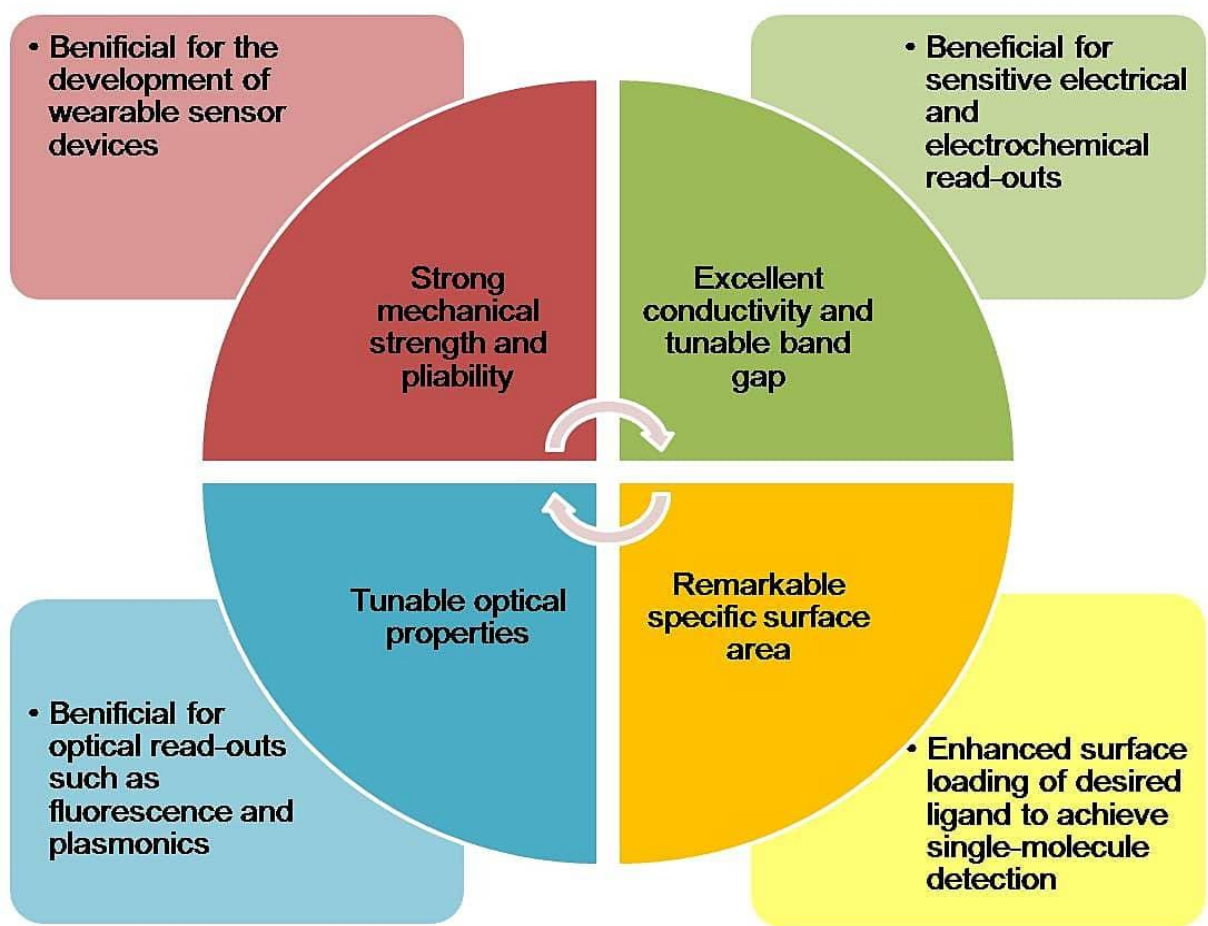

*Fig 2. Novel physical and chemical properties of graphene-based materials facilitate their use in LOC platforms[12].*

The mechanical strength of graphene is very high and it also has the property of pliability. The value of its Young's modulus is 1.0 terapascals with an intrinsic strength of 130 gigapascals, and breaking strength of 42 N/m[13]. The outstanding mechanical strength makes graphene an important material for the fabrication of wearable LOC devices[14]. In addition, graphene possesses excellent electrical conductivity along with a tunable bandgap[15]. Till today, graphene is being considered as the most conductive material [16] having a room-temperature electron mobility of $2 \times 10^5 cm^2 V^{-1} s^{-1}$ with a conductivity of $2.12 \times 10^5$ S/m[17]. These proved beneficial for sensitive electrical and electrochemical read-outs of LOC devices. Graphene also depicts a remarkable specific surface area[18] of 2630 m$^2$/g which helps surface loading of the desired ligand to achieve single-molecule detection in LOC devices. Moreover, graphene exhibits tunable optical properties[19] which are very much advantageous for optical readouts in the case of fluorescence and plasmonic-based LOCs. Graphene also fits well within basic microfluidic fabrication techniques and thus extensively integrated with LOC devices. Thus, graphene along with its derivatives (graphene oxide (GO), reduced graphene oxide (rGO), graphene quantum dots (GQD)) are immensely employed for the fabrication of LOC[20]. Alike most nanomaterials, graphene can be synthesised using both top-down or bottom-up technologies. In the top-down category micromechanical exfoliation, electrochemical exfoliation, thermal exfoliation of graphite intercalation compounds, reduction of GO, arc discharge and unzipping of carbon nanotubes are the techniques that are used for the fabrication of graphene. While sonication, pyrolysis, chemical vapour deposition are considered as the bottom-up methods for graphene synthesis (Fig 3).

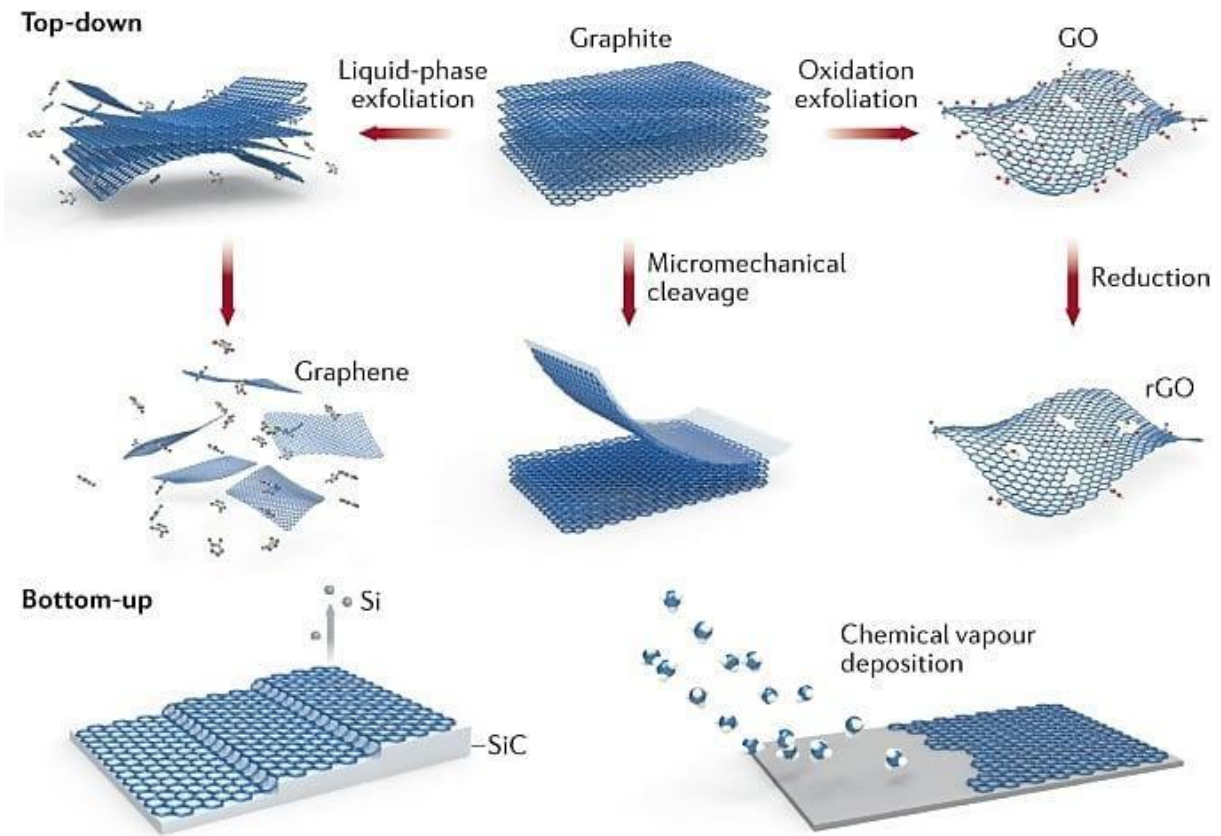

*Fig 3. Various synthesis methods of graphene[21].*

In recent times scientists have devised an advanced graphene synthesis process to curb the cost of production[22,23] up to 95.5%[24] and to achieve it they used ultra-low-cost carbon sources[25]. The low production cost of graphene makes it more preferential for the fabrication of inexpensive LOC platforms with superior performance (Fig 4).

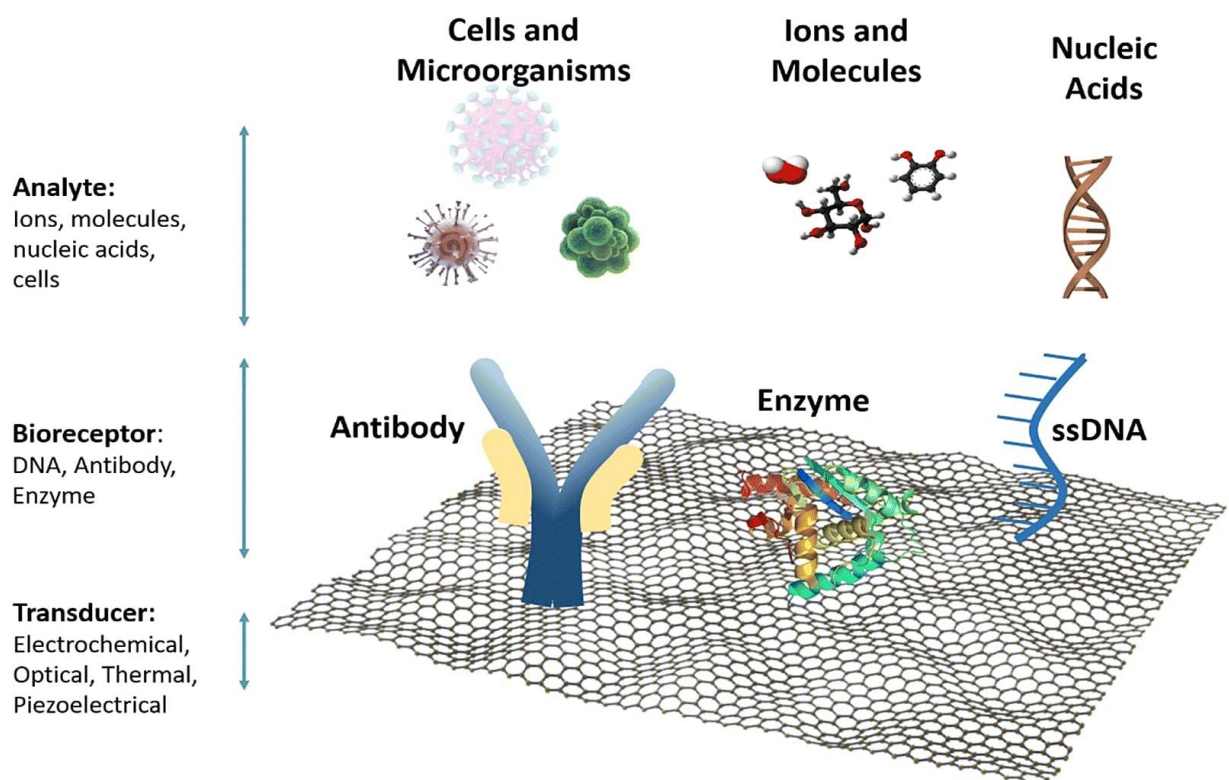

*Fig 4. Various schemes of graphene-based LOC platforms for sensing*[26].

The graphene grown by the above-mentioned methods is utilized in various LOC platforms to perform chemical, biological and point of care analysis. However, only the graphene-based analytical LOC devices used for the detection of viruses are reviewed in this paper.

## 3. Graphene-based lab-on-chip platforms for detection of viruses

Recently, some reviews were published where the use of LOC devices for the detection of only respiratory viruses[27] and the feasibility of using LOC devices for the detection of SARS-CoV-2[28] have been discussed. The graphene-based analytical LOC devices were reported to be employed for the detection of a wide range of viruses' starting from years old Influenza virus to current SARS-CoV-2. Thus, the widest possible range of viruses that are detectable by graphene-based analytical LOC devices is encompassed in this review.

## 3.1 Influenza virus

Nanoparticles of noble elements are widely used along with graphene for the detection of viruses. Huang et al.[29] prepared a sandwich-type immunosensor for the detection of avian influenza virus H7 (AIV H7) employing graphene coated with silver nanoparticles with a lower limit of detection (LOD) of 1.6 pg/mL. Anik et al.[30] employed graphene-gold hybrid nanocomposite to build an electrochemical biosensor for the detection of influenza based on a modified Au-screen printed electrode. The electrochemical route was also used by Veerapandian et al.[31] to fabricate a dual-sensor LOC platform consists of methylene blue-electroadsorbed GO nanostructures modified with antibodies for the detection of influenza A virus at the picomolar level and the response time was extremely small. Chan et al.[32] proposed a unique flow-through microfluidic approach for the detection of influenza virus gene employing rGO based transistor exhibiting LOD of 5 pM with high stability and sensitivity. The microfluidic approach was also adopted by Singh et al.[33] to fabricate rGO based electrochemical immunosensor for the detection of H1N1 with LOD of 0.5 PFU mL$^{-1}$ (Fig 5). The cost of the sensor has always been a major issue to make it market-ready. Working in that direction, Joshi et al.[34] devised a green approach for the fabrication of inexpensive, scalable thermally decomposed rGO flakes for the detection of influenza virus H1N1 with high stability and reproducibility. The functionalization of graphene has provided a new avenue towards biomolecule detection and Roberts et al.[35] used it to develop functionalized graphene-based field-effect transistor (GFET) for the detection of avian influenza virus with LOD 10fM. GFET biosensor structure modified with the sugar chain was employed by Matsumoto et al.[36] for the detection of both the human and avian influenza virus. The same group[37] used a glycan functionalized GFET biosensor for the detection of both the human and avian influenza virus with a LOD of 130 pM and 600 nM, respectively. GFET was also used by this group[38] for the detection of the influenza virus present in the patient's saliva. A Micro-electromechanical system (MEMS) has a wide range of application including biosensor. Chen et al.[39] used this technique to develop a portable GFET biosensor for the detection of the H1N1 virus with a LOD of 1 ng/ml. The fluorometric approach was used by Jeong et al.[40] to fabricate a detection system for influenza virus detection. Multifunctional plasmonic/magnetic graphene was prepared by Lee et al.[41] displayed unique plasmonic and magnetic effects that can be used for the detection of the Influenza virus. The detection technique was highly selective with a lower detection limit of 6.07 pg/ml in human serum. Kinnamon et al.[42] developed a novel screen printed GO-based textile biosensor with high reproducibility and stability. The fabricated biosensor depicted a LOD of 10 ng/mL.

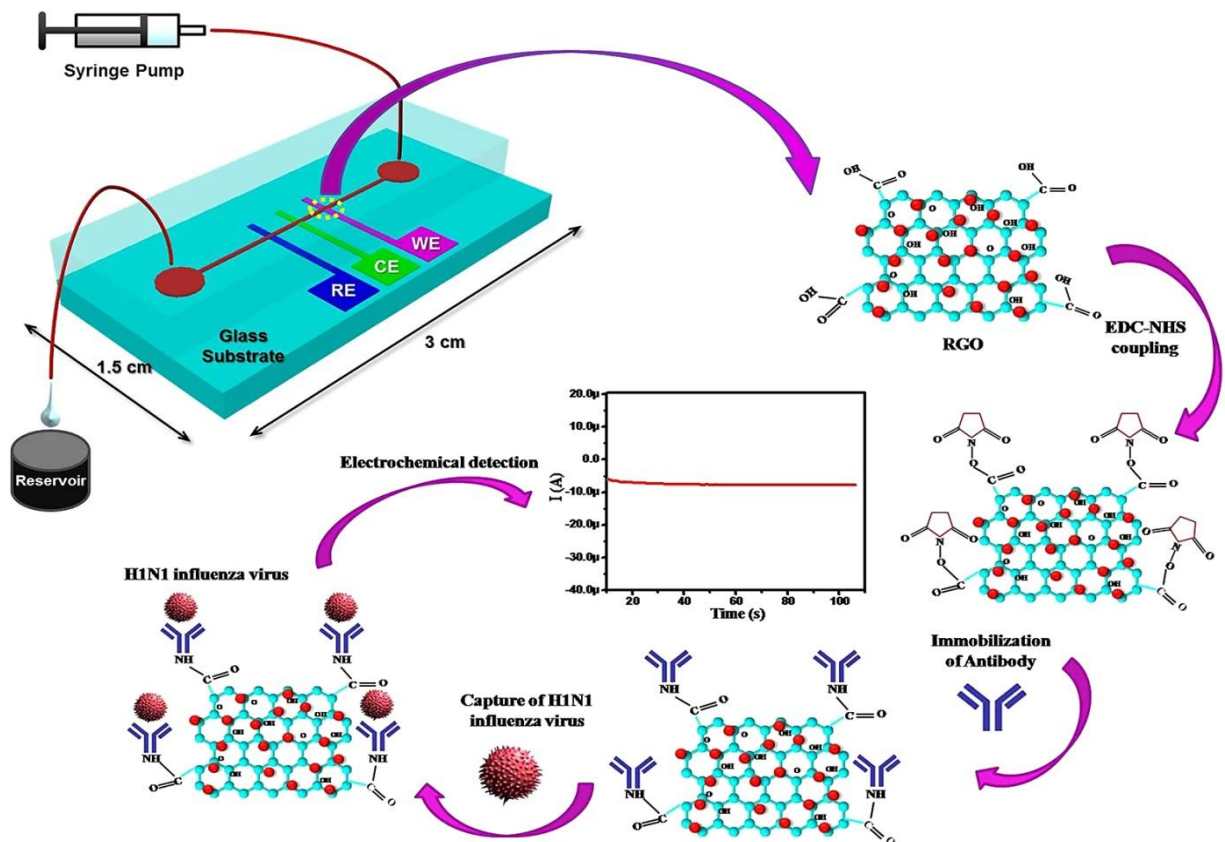

*Fig 5. Schematic illustration of the microfluidics-integrated electrochemical immunosensing chip coated with RGO, followed by antibody immobilization using EDC/NHS coupling for the detection of influenza virus H1N1[43].*

## 3.2 Ebola Virus

Fu et al.[44] used GO to design a supramolecular fluorogenic peptide sensor array for the detection of the Ebola virus. The methods depicted high selectivity in detecting the Ebola virus even in the presence of the Marburg virus and vesicular stomatitis virus. Virus detection using fluorescence quenching property of GO along with rolling circle amplification was employed by Wen et al.[45] to detect the Ebola virus with LOD 1.4 pM both in aqueous solution and 1% serum solution. GFET type biosensors are extensively employed for the detection of the Ebola virus. Maity et al.[46] developed rGO based GFET and used the electronic-resonance frequency, namely inflection frequency to detect the Ebola virus. The sensitivity of the used methodology is several orders of magnitude higher than the sensitivity from other electronic parameters. The Ebola virus was detected by Jin et al.[47] by monitoring the response of the rGO based GFET as a function of the Dirac voltage with LOD of 2.4 pg/mL. Chen et al.[48] developed the rGO based GFET structure and tested the sensor's performance with Ebola virus antigens suspended in $0.01 \times$ PBS/human serum/plasma and the result showed that the device could successfully detect the Ebola virus with the LOD of 1 ng/ml.

## 3.3 Dengue Virus

Jin et al.[49] designed a surface-based impedimetric biosensor using self-assembled functionalized GO wrapped $SiO_2$ particles for label-free detection of dengue virus with LOD 1fM. Opto-electrochemically active ruthenium bipyridine complex was attached to the surface of GO by Kanagavalli et al.[50] for the detection of the dengue virus. The immunosensor showed a linear response both under the chronoamperometric and fluorescence quenching-based immunoassays with a LOD of 0.38 and 0.48 ng/mL, respectively. For the detection of the dengue virus, Omar et al.[51] developed an optical biosensor using surface plasmon resonance of CdS quantum dots composited with amine-functionalized GO thin film, which exhibited a LOD of 1 pM with a sensitivity of 5.49° nM$^{-1}$. They also reported similar biosensing techniques [52,53] using rGO- polyamidoamine nanocomposite which showed a better performance with LOD of 0.08 pM for the detection of dengue virus. The fluorometric approach was adopted by Lee et al.[54] and integrated with a loop-mediated isothermal amplification assay to increase the overall sensitivity of the GO-based biosensor for the detection of dengue with LOD 2.1 nM. Kamil et al.[55] employed a GO integrated biofunctionalized tapered optical fibre based sensor to detect the dengue virus with a LOD of 1 pM along with the sensitivity of 12.77 nm/nM. Electrochemical impedance spectroscopy was employed by Navakul et al.[56] using GO reinforced polymer to detect the dengue virus with a 0.12 pfu/mL detection limit.

### 3.4 SARS-CoV-2 Virus

Graphene was functionalized with SARS-CoV-2 spike antibody by Seo et al.[57] to fabricate a biosensor for the detection of SARS-CoV-2. To immobilize the SARS-CoV-2 spike antibody on GFET they used 1-pyrenebutanoic acid succinimidyl ester as a probe linker. The detection of SARS-CoV-2 using the fabricated device is possible both in the transport medium and clinical sample with a LOD of 1 fg/ml. GFET structure fabricated by Zhang et al.[58] can detect SARS-CoV-2 within 2 mins with a LOD of 0.2 pM. Alafef et al.[59] devised a way for quick, quantitative, inexpensive and ultrasensitive detection of SARS-CoV-2 employing a graphene-based electrochemical biosensor chip containing Au nanoparticles. The performance of the biosensor chip was checked using both SARS-CoV-2 infected Vero cell samples and clinical samples and the chip exhibited a sensitivity of 231 (copies μL$^{-1}$)$^{-1}$ and LOD of 6.9 copies/μL (Fig 6). Simone et al.[60] theoretically studied the possibility of using a graphene-based optical sensor for the detection of SARS-CoV-2 employing Raman scattering. An ultrasensitive supersandwich-type electrochemical sensor was developed by Zhao et al.[61] employing calixarene functionalized GO to detect RNA of SARS-CoV-2. The fabricated device showed a LOD of 200 copies/mL with the clinical specimen moreover, it needs a very small amount of sample (only two copies of SARS-CoV-2) for faithful detection. Hashemi et al.[62] constructed a nanosensor based on the GO-Au nanoparticle to detect the trace of SARS-CoV-2 in any aquatic biological media via the interaction with the active functional groups of their glycoproteins. The device can detect SARS-CoV-2 within 1 min and having the LOD of $1.68 \times 10^{-22}$ µg mL$^{-1}$. Rodríguez et al.[63] devised a unique graphene-based multiplexed immunosensor telemedicine platform employing laser-engraved graphene. The device can perform highly selective, ultrasensitive, and rapid electrochemical detection of SARS-CoV-2 in blood and saliva samples. A paper-based electrochemical immunosensorr for the detection of SARS-CoV-2 was fabricated by Yakoh et al.[64] employing GO. The device can detect SARS-CoV-2 in real clinical sera from COVID patients with LOD of 1 ng/mL and also capable of SARS-CoV-2 antigen detection. Samavati et al.[65] used Au/fiber Bragg grating probe decorated with GO to detect the SARS-CoV-2 virus from patients' saliva and the disease level is identified by the deviation of wavelength and amplitude from healthy saliva. Superfast detection of COVID-19 antibody was made possible by Ali et al.[66] using aerosol jet nanoprinted rGO nanoflakes-coated 3D electrodes integrated with a microfluidic device. The impedance spectroscopy was used to detect the change in impedance upon the presence of SARS-CoV-2 with LOD of $2.8 \times 10^{-15}$ M which can be monitored via a smartphone-based user interface.

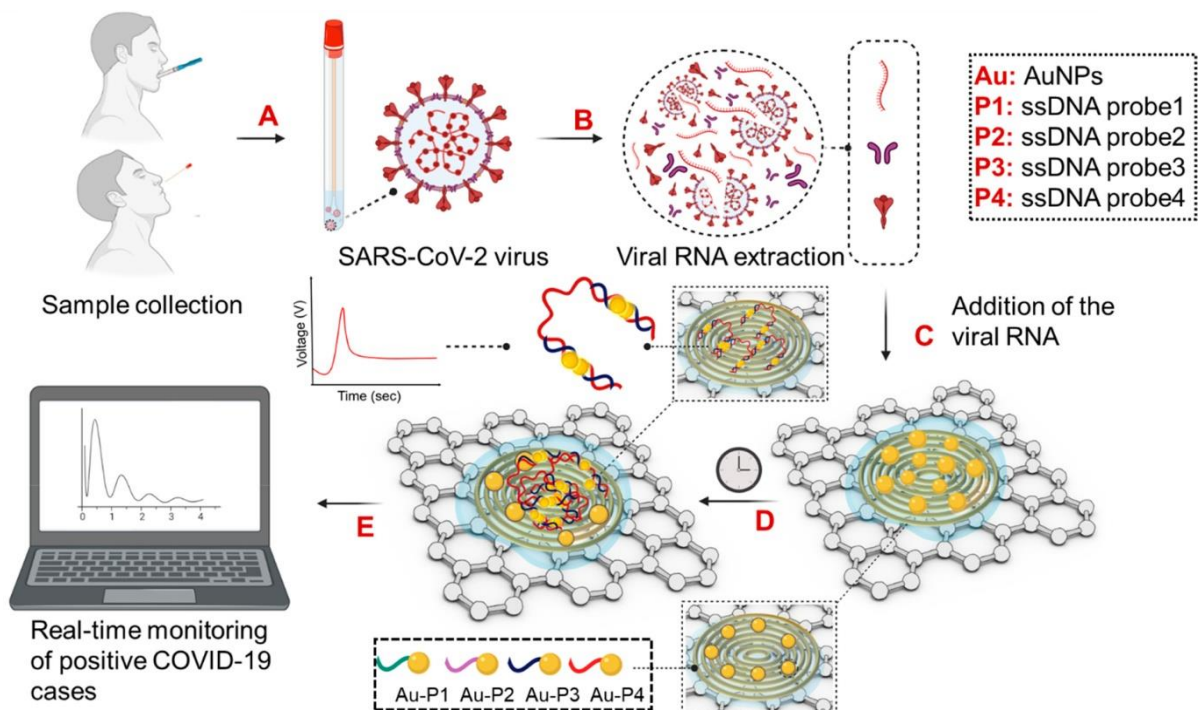

*Fig 6. Scheme 1. Schematic representation of the operation principle of the COVID-19 electrochemical sensing platform wherein step A: the infected samples will be collected from the nasal swab or saliva of the patients under observation; step B: the viral SARS-CoV-2 RNA will be extracted; step C: the viral RNA will be added on top of the graphene-ssDNA-AuNP platform; step D: incubation of 5 min; and step E: the digital electrochemical output will be recorded*[67].

### 3.5 Rotavirus

Liu et al.[68] combined the Hummers method, photolithography and reduction process to fabricate micropatterned rGO FET to detect rotavirus with LOD of $10^2$ pfu. The same group also developed an electrochemical biosensor[69] using GO thin film modified with pyrene derivatives for the detection of rotavirus. Chemi-resistive rGO FET biosensor was developed by Pant et al.[70] and employed pyrene-NHS (Linker) complex to attach the antibodies on the channel of the FET structure for the detection of rotavirus. Fan et al.[71] devised a process for effective detection and removal of rotavirus from infected water using a bioconjugated magnetic nanoparticle-attached hybrid GO-based 3D plasmonic-magnetic solid architecture. Jung et al.[72] developed a GO-based immunobiosensor system to detect norovirus employing fluorescence quenching effect between GO and Au nanoparticle. The maximized GO quenching efficiency was obtained up to 85% at $10^5$ pfu/mL.

**3.6 Hepatitis B Virus**

Gold nanoparticle combined with graphene plays a significant role in the detection of hepatitis B virus. Huang et al.[73] modified graphene electrode with gold nanoparticles and a Nafion-L-cysteine composite film to construct a stable and sensitive electrochemical immunosensor for the detection of hepatitis B with LOD of 0.1 ng mL$^{-1}$. GO-gold nanorod composite was used by Liu et al.[74] for tracing the presence of hepatitis B virus in serum employing surface-enhanced Raman spectroscopy. The fabricated device is specific and highly selective with a LOD of 0.05 pg·mL$^{-1}$. Muain et al.[75] used rGO decorated with gold nanoparticle for impedimetric detection of hepatitis B virus with LOD of 3.80 ng mL$^{-1}$. Graphene resistor, modified with self-assembled graphene-gold nanoparticles was used to sense hepatitis B virus with LOD of 50 pg ml$^{-1}$ by Walters et al.[76]. An electrochemical aptasensor was constructed by Mohsin et al.[77] by modifying the surface of a glassy carbon electrode with gold nanoparticles functionalized rGO. The fabricated device showed an extremely superior LOD of 0.0014 fg/mL. Zhao et al.[78,79] constructed a label-free immunosensor using multilayer film of GO/ferrocene-chitosan nanocomposite and gold nanoparticles for the detection of hepatitis B virus with LOD of 0.01 ng/mL. GQDs are also used for the detection of hepatitis B virus likewise Xiang et al.[80] used GQD based electrochemical biosensing platform that depicts high sensitivity with a LOD of 1 nM. Tan et al.[81] prepared new nanoenzymes using PdCu tripod functionalized porous graphene for the fabrication of hepatitis B biosensor with high selectivity, good stability and reproducibility. The LOD of the constructed biosensor was 20 fg·mL$^{-1}$. GFET structure also finds its way in the successful detection of the hepatitis B virus. Ray et al.[82] used a probabilistic neural network (PNN) for the detection of hepatitis B virus using GFET biosensor. The use of the PNN method showed better performance than its peers and thus lowering the detection limit of the Hepatitis B virus down to 0.1 fM. The nanoporous $SiO_2$ template was employed by Basu et al.[83] to develop an rGO based GFET for sensing the hepatitis B virus to achieve a LOD of 50aM.

**3.7 Hepatitis C Virus**

Kim et al.[84] devised a novel, cost-effective and stable delivery system based on nano GO for simultaneous detection and inhibition of Hepatitis C. The detection was enabled by the DNAzyme-nano GO complex system while DNAzyme mediated catalytic cleavage performed the inhibition of the virus. The silver nanoparticle was employed by Valipour et al.[85] along with GQD to form a nanocomposite platform with a large surface area for the detection of hepatitis C with a LOD of 3 fg mL$^{-1}$. For the quantitative detection of Hepatitis C core antigen, they introduced Riboflavin as a redox probe. Fan et al.[86] employed a fluorescence-based method for the detection of Hepatitis C using rGO nanosheets along with a hybridization chain reaction amplification technique. They studied biological samples under complicated environments and their method exhibited a LOD of 10 fM. Li et al.[87] developed a novel nanocomposite of DNA assisted magnetic rGO-Cu for highly sensitive detection of the Hepatitis C virus. The sensing with a LOD of 405 pM was accomplished using the electrochemical signal originated by $Cu^{2+}$ catalyzed oxidation of O-phenylenediamine. A unique nanocomposite consisting of cucurbit[7]uril and graphene were made by Jiang et al.[88] for electrochemical sensing of Hepatitis C virus with a detection limit as low as 160.4 pmol/L.

**3.8 Papilloma Virus**

Huang et al.[89] employed graphene/Au nanorod/polythionine to construct an ultrasensitive, electrochemical, enzyme-free and label-free DNA biosensor for the detection of human papilloma virus DNA using electrochemical impedance spectroscopy and differential pulse voltammetry. The amplification of the electrochemical signals was achieved by the long-range self-assembly of DNA nanostructure which exhibited a LOD of $4.03 \times 10^{-14}$ mol/L. Chekin et al.[90] used porous rGO/MoS$_2$ to modify glassy carbon electrodes for the sensitive and selective detection of the human papilloma virus employing aptamers as recognition elements. The fabricated electrochemical sensor was able to detect the human papilloma virus with a LOD of 1.75 pM. The electrochemical sensing route was also adopted by Farzin et al.[91] using CNT/amine-ionic liquid functionalized rGO nanoplatforms as a sensing interface for the detection of human papilloma virus in clinical samples with LOD of 13nM. Aptamer-functionalized rGO-FET was constructed by Aspermair et al.[92] to detect the human papilloma virus in saliva and the fabricated device was able to detect the virus with a LOD of 1.75 nM. Mahmoodi et al.[93] prepared a nanocomposite of rGO/MWCNTs and electrodeposited it on a screen-printed carbon electrode. Later on, Au nanoparticles were added and a single-strand DNA probe was immobilized on the modified screen-printed carbon electrode. The differential pulse voltammetry assay was employed for the electrochemical measurement which revealed that the biosensor can detect the virus with a LOD of 0.05 fM.

**3.9 Norovirus**

Graphene decorated with nanoparticles was extensively used in determining the presence of norovirus. Lee et al.[94] prepared a hybrid nanomaterial-sensing platform using $Au/Fe_3O_4$ decorated graphene for sensing norovirus. Owing to the high electrical conductivity and magnetic property of the hybrid structure, the sensing platform showed an exceptional sensitivity level with a LOD of 1.16 pg/ml. Ahmed et al.[95] also prepared an Au-graphene nanohybrid using chloroauric acid, sodium formate and graphene flakes for the detection of norovirus. The catalytic activity of the nanohybrid was employed for the detection of norovirus with a LOD of 92.7 pg/mL. Chand et al.[96] developed an electrochemical aptasensor based on graphene-gold nanocomposite and integrated it with the all-polydimethylsiloxane (PDMS) microfluidic chip. They used differential pulse voltammetric analysis for the sensing of norovirus in spiked blood with LOD of 100pM (Fig 7). A unique biosensing platform was developed by Achadu et al.[97] based on graphene-mediated surface-enhanced Raman scattering (SERS) using plasmonic/magnetic molybdenum trioxide nanocubes for the detection of norovirus. Molybdenum trioxide nanocubes acted as SERS nanotag while graphene acted as the signal reporter. The fabricated biosensor was used to detect norovirus in human fecal samples with a LOD of 5.2 fg/mL. 3D inkjet printed GFET was employed by Xiang et al.[98] as a flexible biosensor platform for the detection of norovirus with the limit of detection around 0.1 µg/ml. Weng et al.[99] adopted an aptamer-based fluorometric approach for the detection of norovirus employing a microfluidic device fabricated using 6-carboxyfluorescein, MWCNT and GO. The hybrid biosensing system was able to detect norovirus with LOD of 3.3 ng·mL$^{-1}$. GO/Fe nanocomposite was prepared by Kim et al.[100] for the detection of norovirus in tap water and artificial urine based on the principle of intra chemiluminescent resonance transfer with LOD of 80 ng/ml.

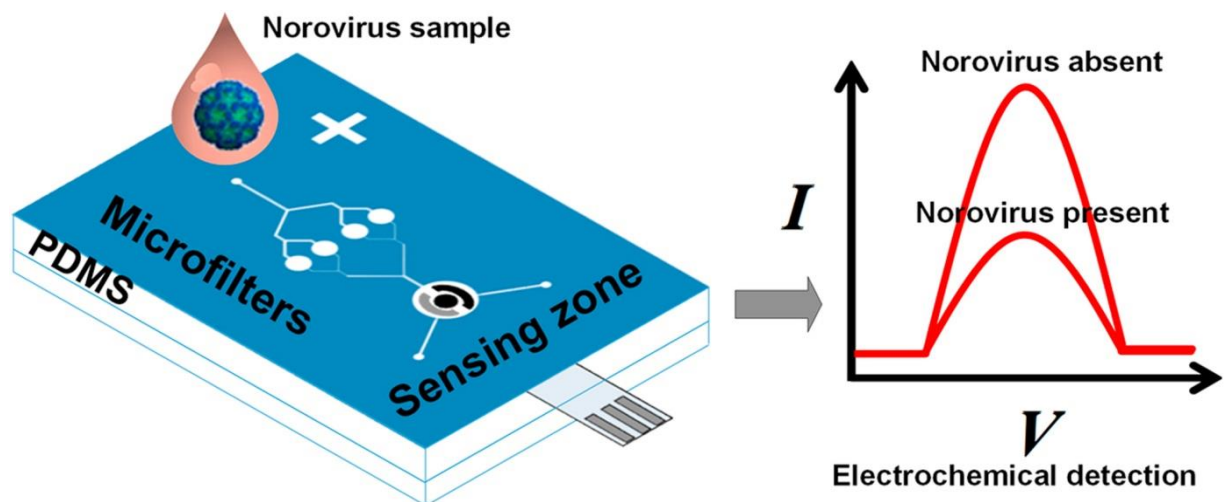

*Fig 7. Microfluidic carbon electrode with Grp-AuNPs composite for Norovirus detection[101].*

### 3.10 Human Immunodeficiency Virus (HIV)

Gong et al.[102] used rGO for the fabrication of impedimetric DNA biosensor to detect HIV with a LOD of $3.0 \times 10^{-13}$ M. Later on, they modified[103] the detection approach by modifying glassy carbon electrode using rGO-nafion composite film for the detection of HIV gene with a LOD of $2.3 \times 10^{-14}$ M. Here stability of graphene coating was enhanced by nafion which increases graphene dispersion. They have also used polyaniline/graphene nanocomposite and dropcasted it onto a glassy carbon electrode for HIV detection[104] and this time the LOD is $1.0 \times 10^{-16}$ M. Detection of HIV using fluorescence polarization was achieved by Wang et al.[105] employing GO as a signal amplifier and the LODof the fabricated chip was 38.6 pmol/L. Kwon et al.[106] used a photolithographic approach to fabricate a scalable and flexible, liquid ion-gated GFET structure and decorated it with close-packed carboxylated polypyrrole nanoparticle arrays to sense the HIV with LOD of 1 pM. Amine-functionalized graphene was employed by Islam et al.[107] to fabricate a GFET biosensor to detect HIV with a LOD of 10 fg/mL. Flexible polyethylene terephthalate was used as a substrate for the fabrication of coplanar-gate GFET by Kim et al.[108] for the detection of HIV. The GFET revealed a remarkable LOD of 47.8 aM in detecting HIV. Vishnubhotla et al.[109] used CVD grown graphene along with the photolithography technique to construct a GFET aptasensor that can detect HIV with LOD of 1 ng/mL. Li et al.[110] studied the employability of fluorescence resonance energy transfer (FRET) technique using boron and nitrogen co-doped GQD to construct a biosensing platform for the detection of HIV and the results exhibited that the device can sense HIV with LOD of 0.5 nM. FRET was also exploited to develop a GO-based sensing platform by Zhang et al.[111] for building a robust and highly sensitive biosensor for the detection of HIV in human serum with a LOD of 1.18 ng/mL. GO-polycarbonate track-etched nanosieve electrochemical biosensing platform was constructed by Nehra et al.[112] for the sensitive and rapid detection of HIV with LOD of 8.3 fM. Wang et al.[113] constructed an electrochemical biosensing platform employing a glassy carbon electrode modified with graphene stabilized gold nanocluster, and integrated with an exonuclease III assisted target recycling amplification strategy to detect HIV with LOD of 30 aM.

### 3.11 Other Viruses

Photolithography along with plasma-enhanced chemical vapour deposition was employed by Afsahi et al.[114] for the fabrication of inexpensive and portable GFET to detect Zika virus in serum which could detect the virus with LOD of 450 pM. Moco et al.[115] opted for the electrochemical route for the fabrication of electrochemical genosensor based on graphite electrodes modified with rGO and polytyramine-conducting polymer which could detect Zika virus with LOD of 0.1 fg/mL within 20 mins. Jin et al.[116] prepared a composite of 3-Aminopropyltriethoxysilane functionalized GO wrapped on $SiO_2$ particles via self-assembly to construct an impedimetric biosensor to detect the Zika virus. An electrochemical immunosensor was constructed by Li et al.[117] using $MoS_2$-rGO nanocomposite decorated with gold nanoparticles to detect the diarrhea virus. Chowdhury et al.[118] employed GQD to construct an electrochemical LOC platform to detect hepatitis E viruses with a LOD of 0.8 fg mL$^{-1}$.

## 4. Challenges and Outlook

The advancement of medical science is not possible without modernization and accelerated clinical diagnosis. Though graphene-based LOC devices extend quick and easy detection of several viruses, yet the recent worldwide outbreak of the SARS-CoV-2 virus not only makes the scientific community realize the challenging aspects of such detection technique but also motivate them in accelerating the research to develop the existing LOC devices with advanced features like integrated-sample preparation, large yield and multiplex detection, high sensitivity or low detection error, integrability with smart devices and finally cost-effectiveness.

Any virus outbreak comes with the challenge of immediate action on impeding further transmission and chain-breaking. An integrated sample preparation capability of existing LOC devices may lead to more rapid detection of the viruses and consequently help in preventing an outbreak. Some recent research showed such incorporation of sample detection capability into LOC[119,120]. However, consolidated graphene-based LOC with single-chip sample preparation or extraction is still lacking. Whether it is a home-based or a clinical centre-based diagnosis, futuristic automation, as well as rapid identification of viral infection, will not be possible without the advent of the LOC device. Integrated sample preparation within LOC will enable rapid detection of the infection with minimal delay and thus the patients can be offered immediate treatment to prevent a fatality.

The recent outbreak of SARS-CoV-2 exerts a thrust into clinical research highlighting several critical requirements like the multi-detection capability of LOC devices. A single viral infection often comes with multi-variant infection in human bodies leading to even loss of human life without early detection. This type of situation in medical science demands highly sensitive and multiplex detection. Consequently, multi-variant infection detection through a single chip is of utmost importance for the recent advancement of LOC devices. Moreover, in most cases, viral infections having numerous sub-types are found to spread quickly and impose a crucial challenge for LOC devices. To cater greater portion symptomatic population through a single detection chip, the graphene-based LOC devices are required to have high throughput as evidenced in the case of some of the microfluidic chips in the published literature[121,122]. Graphene-based LOC with high throughput and multiplex detection capability can effectively decrease the cost-per-patient criterion and thereby can widen its use.

Some recent studies[123,124] reported the efficacy of the digital quantitative method in terms of its better precision of detection in comparison to the standard quantitative method. Advanced technology like robotic process automation can be clubbed to the existing clinical LOC devices to realize "sample-in-digital-answer-out". Nevertheless, such technology presently associated with certain limitations like cost and implementation. To cope-up with the futuristic technological advancement, the graphene-based LOCs should confront the challenge of the digital quantitative method through more rigorous multi-disciplinary research activities.

In the case of infectious diseases, the clinical diagnosis of patients with low viral load always faces a serious challenge to existing technologies. Among the available techniques, plasmonic and electrochemical probing of the clinical detection process possesses the highest sensitivity in viral load detection. Moreover, the plasmonic investigation inherits advantages like highest throughput as well as fastest read-out among other significant assays like colorimetric, lateral flow, magnetic, and electrochemical approaches. More rigorous scientific investigation can lead to an overall improved version of LOC.

Because of the present technological evolution, the integration of existing LOC detection with smart devices is becoming an utmost requisite in modern state-of-the-art living. Recent studies had already indicated new technologies to enable the LOC to function in conjunction with smart read-out capabilities[125].

Finally, all the above-mentioned features of the LOC unit are required to be implemented in a cost-effective approach because of the large-scale spreading aspect of a virus outbreak that can even lead to another pandemic scenario without immediate attention (Fig 8).

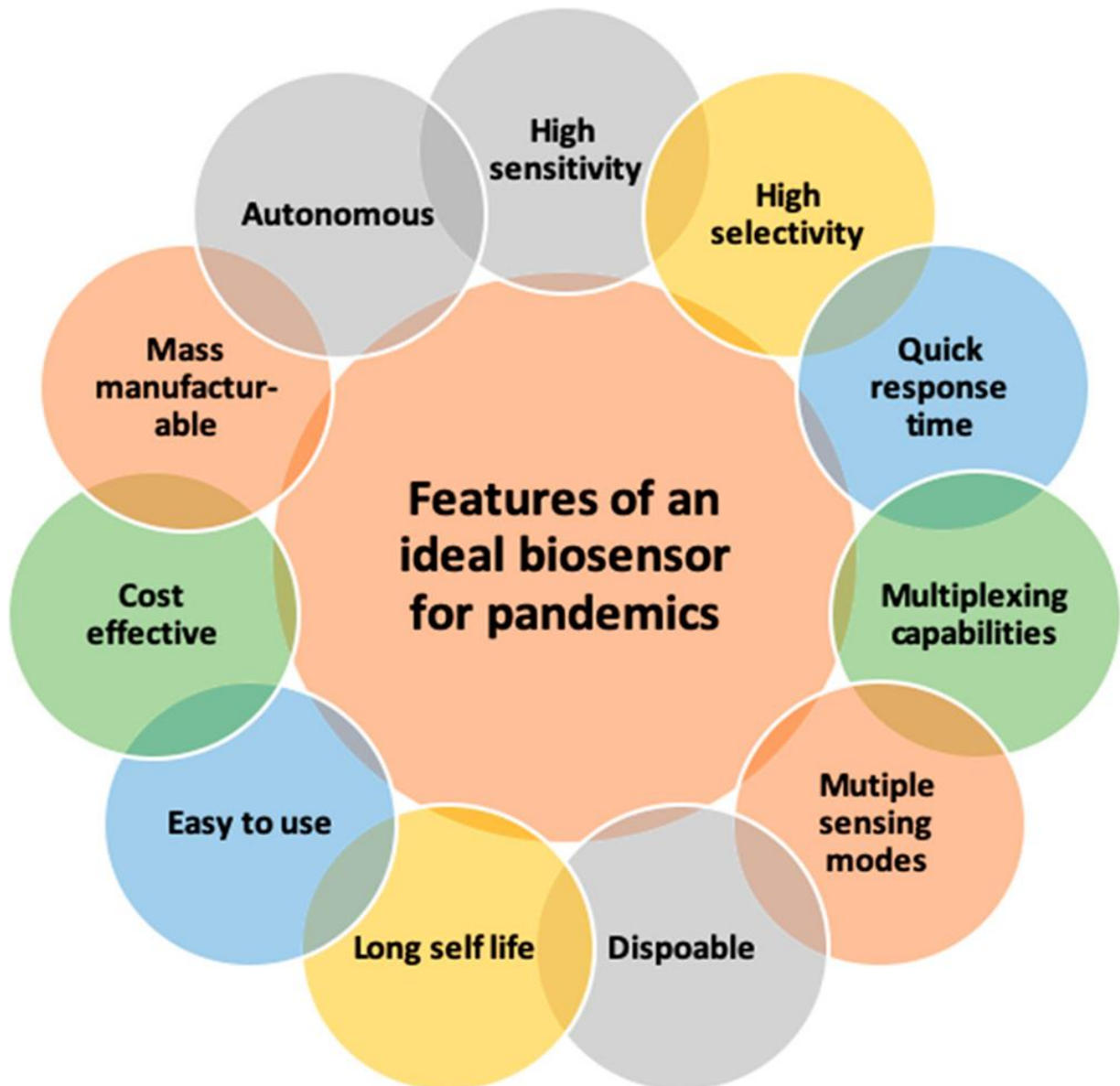

*Fig 8. Features of an ideal biosensor required to be developed for effective use[126].*

**5. Conclusions**

Graphene-based LOC devices offer the advantages of novel physiochemical properties of graphene along with exotic analytical characteristics of microfluidics namely minimal consumption of target molecules, rapid detection, user-friendliness with high sensitivity. Here, firstly the fabrication process technologies for LOC device has been discussed and thereafter advantages of using graphene in LOC devices have been explored. Later on, the application of graphene-based LOC in detecting a wide spectrum of viruses has been reviewed which ranges from the detection of years old influenza virus to modern SARS-CoV-2. From the economic perspective, graphene depicts a huge potential as recent developments show a steep decrease in the production cost of high-quality graphene. However, still, there are some issues namely toxicity, shelf life and sustainability, which are needed to be addressed with utmost concern. Reproducibility plays a major role in any type of nanodevice and graphene-based LOC comes under it. Moreover, as graphene-based LOCs are mostly meant for point of care related application, thus the process integrity and material reproducibility need an extensive quality check. In summary, till now graphene along with its derivatives has overcome most of the hurdles, thus it can be believed that an eco-friendly, inexpensive, reproducible and sustainable graphene-based LOC platform will pave a new way for the detection of almost all kinds of viruses.

[77] Dheyaa Hussein Mohsin, Muthana Saleh Mashkour, and Fataneh Fatemi, 'Design of Aptamer-Based Sensing Platform Using Gold Nanoparticles Functionalized Reduced Graphene Oxide for Ultrasensitive Detection of Hepatitis B Virus', *Chemical Papers*, 4 August 2020, https://doi.org/10.1007/s11696-020-01292-1.

[78] Feijun Zhao et al., 'Label-Free Amperometric Immunosensor Based on Graphene Oxide and Ferrocene-Chitosan Nanocomposites for Detection of Hepatis B Virus Antigen', *Journal of Biomedical Nanotechnology* 13, no. 10 (1 October 2017): 1300–1308, https://doi.org/10.1166/jbn.2017.2415.

[79] Feijun Zhao et al., 'An Electrochemical Immunosensor with Graphene-Oxide-Ferrocene-Based Nanocomposites for Hepatitis B Surface Antigen Detection', *Electroanalysis* 30, no. 11 (2018): 2774–80, https://doi.org/10.1002/elan.201800476.

[80] Qian Xiang et al., 'A Label-Free Electrochemical Platform for the Highly Sensitive Detection of Hepatitis B Virus DNA Using Graphene Quantum Dots', *RSC Advances* 8, no. 4 (2018): 1820–25, https://doi.org/10.1039/C7RA11945C.

[81] Zhaoling Tan et al., 'A Label-Free Immunosensor for the Sensitive Detection of Hepatitis B e Antigen Based on PdCu Tripod Functionalized Porous Graphene Nanoenzymes', *Bioelectrochemistry* 133 (1 June 2020): 107461, https://doi.org/10.1016/j.bioelechem.2020.107461.

[82] R. Ray et al., 'Label-Free Biomolecule Detection in Physiological Solutions With Enhanced Sensitivity Using Graphene Nanogrids FET Biosensor', *IEEE Transactions on NanoBioscience* 17, no. 4 (October 2018): 433–42, https://doi.org/10.1109/TNB.2018.2863734.

[83] J. Basu and C. RoyChaudhuri, 'Attomolar Sensitivity of FET Biosensor Based on Smooth and Reliable Graphene Nanogrids', *IEEE Electron Device Letters* 37, no. 4 (April 2016): 492–95, https://doi.org/10.1109/LED.2016.2526064.

[84] Seongchan Kim et al., 'Deoxyribozyme-Loaded Nano-Graphene Oxide for Simultaneous Sensing and Silencing of the Hepatitis C Virus Gene in Liver Cells', *Chemical Communications* 49, no. 74 (20 August 2013): 8241–43, https://doi.org/10.1039/C3CC43368D.

[85] Akram Valipour and Mahmoud Roushani, 'Using Silver Nanoparticle and Thiol Graphene Quantum Dots Nanocomposite as a Substratum to Load Antibody for Detection of Hepatitis C Virus Core Antigen: Electrochemical Oxidation of Riboflavin Was Used as Redox Probe', *Biosensors and Bioelectronics* 89 (15 March 2017): 946–51, https://doi.org/10.1016/j.bios.2016.09.086.

[86] Jialong Fan et al., 'An Ultrasensitive and Simple Assay for the Hepatitis C Virus Using a Reduced Graphene Oxide-Assisted Hybridization Chain Reaction', *Analyst* 144, no. 13 (24 June 2019): 3972–79, https://doi.org/10.1039/C9AN00179D.

[87] Jingwen Li et al., 'Sensitive Electrochemical Detection of Hepatitis C Virus Subtype Based on Nucleotides Assisted Magnetic Reduced Graphene Oxide-Copper Nano-Composite', *Electrochemistry Communications* 110 (1 January 2020): 106601, https://doi.org/10.1016/j.elecom.2019.106601.

[88] Ping Jiang et al., 'Ultrasensitive Detection of Hepatitis C Virus DNA Subtypes Based on Cucurbituril and Graphene Oxide Nano-Composite', *Chemical Research in Chinese Universities* 36, no. 2 (1 April 2020): 307–12, https://doi.org/10.1007/s40242-020-9111-8.

[89] Huayu Huang et al., 'An Ultrasensitive Electrochemical DNA Biosensor Based on Graphene/Au Nanorod/Polythionine for Human Papillomavirus DNA Detection', *Biosensors and Bioelectronics* 68 (15 June 2015): 442–46, https://doi.org/10.1016/j.bios.2015.01.039.

[90] Fereshteh Chekin et al., 'Nucleic Aptamer Modified Porous Reduced Graphene Oxide/MoS2 Based Electrodes for Viral Detection: Application to Human Papillomavirus (HPV)', *Sensors and Actuators B: Chemical* 262 (1 June 2018): 991–1000, https://doi.org/10.1016/j.snb.2018.02.065.

[91] Leila Farzin et al., 'Electrochemical Genosensor Based on Carbon Nanotube/Amine-Ionic Liquid Functionalized Reduced Graphene Oxide Nanoplatform for Detection of Human Papillomavirus (HPV16)-Related Head and Neck Cancer', *Journal of Pharmaceutical and Biomedical Analysis* 179 (5 February 2020): 112989, https://doi.org/10.1016/j.jpba.2019.112989.

[92] Patrik Aspermair et al., 'Reduced Graphene Oxide–Based Field Effect Transistors for the Detection of E7 Protein of Human Papillomavirus in Saliva', *Analytical and Bioanalytical Chemistry*, 20 August 2020, https://doi.org/10.1007/s00216-020-02879-z.

[93] Pegah Mahmoodi et al., 'Early-Stage Cervical Cancer Diagnosis Based on an Ultra-Sensitive Electrochemical DNA Nanobiosensor for HPV-18 Detection in Real Samples', *Journal of Nanobiotechnology* 18, no. 1 (13 January 2020): 11, https://doi.org/10.1186/s12951-020-0577-9.

[94] Jaewook Lee et al., 'Binary Nanoparticle Graphene Hybrid Structure-Based Highly Sensitive Biosensing Platform for Norovirus-Like Particle Detection', *ACS Applied Materials & Interfaces* 9, no. 32 (16 August 2017): 27298–304, https://doi.org/10.1021/acsami.7b07012.